\documentclass[prd,twocolumn,showpacs,amsmath,amssymb]{revtex4}
\usepackage{graphicx}
\usepackage{dcolumn}
\usepackage{bm}

\newcommand{\Dp}{D^{\prime}}
\newcommand{\Fp}{F^{\prime}}

\newcommand{\ag}{a_{3g}}
\newcommand{\aga}{a_{\gamma}}
\newcommand{\ac}{a_c}

\newcommand{\pip}{\pi^+}
\newcommand{\pim}{\pi^-}
\newcommand{\piz}{\pi^0}
\newcommand{\kap}{K^+}
\newcommand{\kam}{K^-}
\newcommand{\kaz}{K^0}
\newcommand{\kazb}{\overline{K}^0}

\newcommand{\kstp}{K^{*+}}
\newcommand{\kstm}{K^{*-}}
\newcommand{\kstz}{K^{*0}}
\newcommand{\kstzb}{\overline{K}^{*0}}

\newcommand{\rop}{\rho^+}
\newcommand{\rom}{\rho^-}
\newcommand{\roz}{\rho^0}

\newcommand{\psp}{\psi^{\prime}}

\newcommand{\jpsi}{J/\psi}

\newcommand{\EE}{e^+e^-}
\newcommand{\MM}{\mu^+\mu^-}

\newcommand{\pp}{\pi^+\pi^-}

\newcommand{\kskl}{K^0_S K^0_L}

\newcommand{\ccbar}{c\bar{c}}

\newcommand{\nnb}{n\overline{n}}

\newcommand{\ppb}{p\overline{p}}

\newcommand{\LLb}{\Lambda \overline{\Lambda}}

\newcommand{\SSbz}{\Sigma^0 \overline{\Sigma}^0}
\newcommand{\SSbp}{\Sigma^+ \overline{\Sigma}^-}
\newcommand{\SSbm}{\Sigma^- \overline{\Sigma}^+}
\newcommand{\XXbz}{\Xi^0 \overline{\Xi}^0}
\newcommand{\XXbm}{\Xi^- \overline{\Xi}^+}
\newcommand{\SzLb}{\Sigma^0 \overline{\Lambda}}

\newcommand{\SbzL}{\overline{\Sigma}^0 \Lambda}
\newcommand{\DDltpp}{\Delta^{++} \overline{\Delta}^{--}}
\newcommand{\DDltp}{\Delta^{+} \overline{\Delta}^{-}}
\newcommand{\DDltz}{\Delta^{0} \overline{\Delta}^{0}}
\newcommand{\DDltn}{\Delta^{-} \overline{\Delta}^{+}}
\newcommand{\OOb}{\Omega^{-} \overline{\Omega}^{+}}

\newcommand{\Heff}{{\cal H}_{eff}}
\newcommand{\Hz}{H_{0}}
\newcommand{\Hmass}{H_{3}^{3}}
\newcommand{\Hchrg}{H_{1}^{1}}
\newcommand{\gz}{g_{0}}
\newcommand{\bspaf}{\stackrel{\leftrightarrow}{\partial}}

\newcommand{\beq}{\begin{equation}}
\newcommand{\eeq}{\end{equation}}
\newcommand{\beqn}{\begin{eqnarray}}
\newcommand{\eeqn}{\end{eqnarray}}
\newcommand{\beqns}{\begin{eqnarray*}}
\newcommand{\eeqns}{\end{eqnarray*}}
\newcommand{\bfg}{\begin{figure}}
\newcommand{\efg}{\end{figure}}
\newcommand{\bitm}{\begin{itemize}}
\newcommand{\eitm}{\end{itemize}}
\newcommand{\bnum}{\begin{enumerate}}
\newcommand{\enum}{\end{enumerate}}
\newcommand{\btbl}{\begin{table}}
\newcommand{\etbl}{\end{table}}
\newcommand{\btbu}{\begin{tabular}}
\newcommand{\etbu}{\end{tabular}}
\def\eref#1{(\ref{#1})}
\def\Journal#1#2#3#4{{#1} {\bf #2}, #3 (#4)}

\def\PLB{Phys. Lett. B}

\def\PRL{Phys. Rev. Lett.}
\def\PRD{Phys. Rev. D}

\def\HEPNP{HEP \& NP}

\def\prd#1#2#3 {{~Phys. Rev. D {#1}, #2 (#3) }}  
\def\plb#1#2#3 {{~Phys. Lett. B {#1}, #2 (#3) }}  

\newsavebox{\arrect}
\savebox{\arrect}(0,0){%
\setlength{\unitlength}{0.5cm}
\thicklines
\put(-4,1){\line(1,0){8.0}}
\put(4,1){\line(0,-1){2.0}}
\put(4,-1){\line(-1,0){8.0}}
\put(-4,-1){\line(0,1){2.0}}
\put(0,-1){\vector(0,-1){1.0}}
}

\newsavebox{\arrhomb}
\savebox{\arrhomb}(0,0){%
\setlength{\unitlength}{1cm}
\thicklines
\put(0,0.5){\line(-2,-1){1.0}}
\put(0,0.5){\line(2,-1){1.0}}
\put(0,-0.5){\line(-2,1){1.0}}
\put(0,-0.5){\line(2,1){1.0}}
\put(0,-0.5){\vector(0,-1){0.5}}
\put(-0.15,-0.7){\makebox(0,0)[r]{no}}
\put(1.0,0.0){\vector(1,0){2.0}}
\put(2.0,0.05){\makebox(0,0)[b]{yes}}
}

\newsavebox{\arrparall}
\savebox{\arrparall}(0,0){%
\setlength{\unitlength}{0.5cm}
\thicklines
\put(-3,1){\line(1,0){8.0}}
\put(5,1){\line(-1,-1){2.0}}
\put(3,-1){\line(-1,0){8.0}}
\put(-5,-1){\line(1,1){2.0}}
\put(0,-1){\vector(0,-1){1.0}}
}
\newsavebox{\arrparalla}
\savebox{\arrparalla}(0,0){%
\setlength{\unitlength}{0.5cm}
\thicklines
\put(-3,1){\line(1,0){8.0}}
\put(5,1){\line(-1,-1){2.0}}
\put(3,-1){\line(-1,0){8.0}}
\put(-5,-1){\line(1,1){2.0}}
}

\begin{document}

\title{Symmetry analysis of charmonium two-body decay}

\author{X.~H.~Mo$^{1,2}$, P.~Wang$^{1}$, J.~Y.~Zhang$^{1}$
\\  \vspace{0.2cm} {\it
$^{1}$ Institute of High Energy Physics, CAS, Beijing 100049, China\\
$^{2}$ University of Chinese Academy of Sciences, Beijing 100049, China\\
}
}
\email{moxh@ihep.ac.cn}

\date{\today}
	
\begin{abstract}
In the light of $SU(3)$ flavor symmetry, the effective interaction Hamiltonian in tensor form is obtained by virtue of group representation theory. The strong and electromagnetic breaking effects are treated as a spurion octet so that the flavor singlet principle can be utilized as the criterion to determine the form of effective Hamiltonian. Two body decays of both baryonic and mesonic final states are parameterized in the uniform scheme, based on which the relative phase between the strong and electromagnetic amplitudes is studied for various charmonium decay modes, including $\psp$ and/or $\jpsi$ decay to octet baryon pair, decuplet baryon pair, decuplet-octet baryon final state, and pseudoscalar-pseudoscalar meson final state. In data analysis of samples taken in $\EE$ collider, the details of experimental effects, such as energy spread and initial state radiative correction are taken into consideration in order to make full use of experimental information and acquire the accurate and delicate results.
\end{abstract}
\maketitle

\section{Introduction}
The Standard Model (SM) has been accepted as a universally appreciated theory basis in high energy community, and mainly consists of two parts. One is Salam-Weinberg model that depicts the electro-weak interaction, which can usually accommodates accurate enough evaluation for certain process. Another part of SM is quantum chromodynamics (QCD) that depicts the strong interaction. It has been proved to be very successful at high energy when the calculation can be executed perturbatively. Nevertheless, its validity at non-perturbative regime needs more experimental guidance. The production and decay of charmonium states supply an ideal laboratory for such a study.

Charmonium is the bound state of a charm quark and an anti-charm quark, it is one of the simplest system bound by strong force. As the mass of the charmonium states is between 3 and 5 GeV, the transition regime between the perturbative and non-perturbative QCD, so it is extremely important in both theory and experiment. One may remember that at the early stage of the discovery of a narrow state $\jpsi$, the $\ccbar$ system was hailed as the Hydrogen atom of QCD, with the implied hope that the study of the newly discovered system could shed as much light on the dynamics of quark-antiquark interactions as the study of the Hydrogen atom had on Quantum Electrodynamics. But one may also notice the historical fact, even before Bohr's theory, Balmer series had been discovered for long, and the famous Rydberg formula had also been proposed, which laid a solid foundation for further theoretical improvement. If we are conscious of the more complicatedness of charmonia system comparing with the Hydrogen atom, we may prepare for more hard and meticulous works.

As one of important and interesting steps, it is a good start point to study the relative phase between the strong and electromagnetic (EM) interaction amplitudes, which provides us a new viewpoint to explore the quarkonium decay dynamics, then profound our understanding on QCD. Studies have been carried out for many $\jpsi$ and $\psp$ two-body mesonic decay modes with various spin-parities: $1^-0^-$~\cite{dm2exp,mk3exp,wymphase}, $0^-0^-$~\cite{a00,LopezCastro:1994xw,wymppdk,a11}, and $1^-1^-$~\cite{a11}, and baryon antibaryon pairs~\cite{ann}. These analyses reveal that there exists a relative orthogonal phase between the EM and strong decay amplitudes~\cite{dm2exp,mk3exp,wymphase,a00,LopezCastro:1994xw,wymppdk,a11,ann,suzuki}. There is also a conjecture to claim that such an orthogonal phase is universal for all quarkonia decays~\cite{Wang:2003zx}.

Besides the experimental measurement of the phase, there is also some theoretical efforts to parameterize the various decay modes to provide us more insight of decay dynamics~\cite{Kowalski:1976mc,Clavelli:1983,ssPinsky,Haber,Seiden88,Morisita:1990cg,zmy2015,Baldini19,moxh2022}, such as the pseudoscalar and pseudoscalar mesons (PP), vector and pseudoscalar mesons (VP), octet baryon-pair, and so on. In this monograph, based on the concise flavor singlet principle, two body decay modes, including both mesonic and baryonic final states, are parameterized systematically and consistently. And this kind of results definitely facilitates the study of the relative phase between EM and strong interactions.

In next section, the parametrization scheme will be expounded firstly, then the effective Hamiltonians for both baryonic and mesonic decay modes are obtained consecutively. Section that follows discusses in detail the special experiment effects of $\EE$ collider, then the successive section focuses on concrete data analysis. The last section is a summary.

\section{Analysis framework}\label{xct_alsfrk}
In $\EE$ collider experiment, the initial state is obviously flavorless, then the final state must be flavor singlet. Moreover, only the Okubo-Zweig-Iizuka (OZI) rule suppressed processes are considered and the final states merely involve light quarks, that is $u, d, s$ quarks. Therefore, solely the theory of unitary $SU(3)$ group is employed for symmetry analysis. The key rule herein is the so-called ``flavor singlet principle'' that determines what kinds of terms are permitted in effective interaction Hamiltonian. Resort to the perturbation language, the Hamiltonian is written as
\beq
\Heff = H_0 + \Delta H~,
\label{perturbaionhmtn}
\eeq
where $H_0$ is the symmetry conserved term and $\Delta H$ the symmetry breaking term, which is generally small compare to $H_0$. Since we focus on two-body decay, merely two multiplets, say ${\mathbf n}$ and ${\mathbf m}$, need to be considered. In the light of group representation theory, the product of two multiplets can be decomposed into a series of irreducible representations, that is
\beq
{\mathbf n} \otimes {\mathbf m} = {\mathbf l_1} \oplus {\mathbf l_2} \oplus \cdots \oplus {\mathbf
l_k}~.
\label{dcpsmoftwomtplt}
\eeq
The singlet principle requires that among the ${\mathbf l_j} (j=1, \cdots, k)$, only the singlet term, i.e. ${\mathbf l_j}={\mathbf 1}$ for certain $j$, can be allowed in the Hamiltonian. Since this term is obviously $SU(3)$ invariant, it is called the symmetry conserved term, i.e. $H_0$.

Now turn to $SU(3)$-breaking term. Two types of $SU(3)$ breaking effect are to be considered.  One is the mass breaking term. Here, $SU(2)$ isospin symmetry (or $I$-spin symmetry in group language) is assumed, that is $m_u=m_d$; but $m_s \neq m_u, m_d$ and this mass difference between $s$ and $u$/$d$ quarks leads to $SU(3)$ breaking. In $SU(3)$ fundamental representation, with Gell-Mann matrices, such a mass breaking effect can be described by matrix ${\mathbf S}_{m}$,
\beq
 {\mathbf S}_{m}= \frac{g_m}{3} \left(\begin{array}{ccc}
1 &   &  \\
  & 1 &  \\
  &   & -2
\end{array}\right)~,
\label{smbrkmass}
\eeq
where $g_m$ is the effective coupling constant due to mass difference effect.

Along the same line, electromagnetic effect also violates $SU(3)$ invariance but keeps charge symmetry (or $U$-spin symmetry in group language), such a charge breaking effect is described by matrix ${\mathbf S}_{e}$,
\beq
{\mathbf S}_{e}= \frac{g_e}{3}
\left(\begin{array}{ccc}
2 &    &  \\
  & -1 &  \\
  &    & -1
\end{array}\right)~,
\label{smbrkcharge}
 \eeq
where $g_e$ is the effective coupling constant due to charge difference effect.

It is well-known that octet hadron, meson and/or baryon can be expressed by Gell-Mann matrices as well. By virtue of Eqs.~\eref{smbrkmass} and \eref{smbrkcharge}, it inspires us to consider the $SU(3)$-breaking as one kind of octet. Following the recipe proposed in Ref.~\cite{Haber}, this kind of $SU(3)$-breaking effect is called a ``spurion'' octet. With this notion, in order to figure out the breaking term in the Hamiltonian, the products of this spurion octet with the irreducible representations ${\mathbf l_j} (j=1, \cdots, k)$ will be scrutinized, only the singlet term in the decomposition will be allowed in the Hamiltonian. Concretely,
\beq
{\mathbf l_j} \otimes {\mathbf 8} = {\mathbf q_1} \oplus {\mathbf q_2} \oplus \cdots \oplus {\mathbf
q_k}~,
\label{dcpsmofoctandirrds}
\eeq
then if and only if some ${\mathbf q_i}=1$, the corresponding term is allowed. Since such a kind of term violates $SU(3)$ invariance, it is called the symmetry breaking term. In a word, with the singlet principle, the effective interaction Hamiltonian can be determined definitely.

Now there are two issues need to be explained. Firstly, by virtue of group representation theory, the particles can be expressed in many notations, here the tensor denotation is adopted, so that all multiplets can be expressed consistently. Under this form, how to express the breaking term? In the light of Eqs.~\eref{smbrkmass} and \eref{smbrkcharge}, it is noticed that ${\mathbf S}_{m}$ is $I$-spin conserved breaking while ${\mathbf S}_{e}$ $U$-spin conserved breaking, this is equivalent to contract the superscript and subscript indexes along 3 and 3 direction to obtained the mass breaking term $\Hmass$, and contract the superscript and subscript indexes along 1 and 1 direction to obtained the charge breaking term $\Hchrg$.

Secondly, the Hamiltonian term is usually written as $\psi M_1 M_2$, where $\psi$ indicates the charmonium state while $M_1$ and $M_2$ are two multiplet components with the corresponding tensor indexes contracted. Since the charmonium state $\psi$ is the same for whole final state, and what we care about is the relative relation of multiplet, so $\psi$ is suppressed in the following derivations.

\subsection{Parametrization of baryonic final state}
We start with decuplet-decuplet baryon pair final state. In $SU(3)$ classification, the decuplet contains the isospin multiplets $I=0,\frac{1}{2},1,$ and $\frac{3}{2}$ corresponding respectively to the tensor components $B^{333},B^{i33},B^{ij3},$ and $B^{ijk}$, for $i,j,k=1,2$. These are assigned to the lowest excited baryon states~\cite{quangpham}:
\beq
\left.\begin{array}{llll}
B^{111}=\Delta^{++}&B^{112}=\frac{1}{\sqrt{3}}\Delta^{+}&B^{122}=\frac{1}{\sqrt{3}}\Delta^{0}&B^{222}=\Delta^{-}\\
B^{113}=\frac{1}{\sqrt{3}}\Sigma^{+}&B^{123}=\frac{1}{\sqrt{6}}\Sigma^{0}& B^{223}=\frac{1}{\sqrt{3}}\Sigma^{-}&     \\
B^{133}=\frac{1}{\sqrt{3}}\Xi^{0}&  B^{233}=\frac{1}{\sqrt{3}}\Xi^{-}&  &    \\
B^{333}=\Omega^-     &             &
\end{array}\right.
\label{dkpbyn}
\eeq
The related anti-baryon is denoted as $B_{ijk}$, that is $B_{ijk}=\overline{B}^{ijk}$~:
\beq
\left.\begin{array}{llll}
B_{111}=\overline{\Delta}^{--}
   &B_{112}=\frac{1}{\sqrt{3}}\overline{\Delta}^{-}
      &B_{122}=\frac{1}{\sqrt{3}}\overline{\Delta}^{0}
         &B_{222}=\overline{\Delta}^{+}\\
B_{113}=\frac{1}{\sqrt{3}}\overline{\Sigma}^{-}
   &B_{123}=\frac{1}{\sqrt{6}}\overline{\Sigma}^{0}
      & B_{223}=\frac{1}{\sqrt{3}}\overline{\Sigma}^{+}&     \\
B_{133}=\frac{1}{\sqrt{3}}\overline{\Xi}^{0}
      & B_{233}=\frac{1}{\sqrt{3}}\overline{\Xi}^{+}&  &    \\
B_{333}=\overline{\Omega}^+     &             &
\end{array}\right.
\label{dkpabyn}
\eeq
It also should be noted that $\Sigma$ and $\Xi$ in decuplet are conventionally denoted as ${\Sigma^*}$ and ${\Xi^*}$ to indicate the excited states, but the star in superscript is suppressed in this subsection without ambiguity. However, when discussing the decuplet-octet final state, the symbol will be recovered to avoid confusion.

According to group theory, the product of two decuplets can be decomposed as follows
\beq
{\mathbf 10} \otimes {\mathbf 10^*} = {\mathbf 1} \oplus {\mathbf 8} \oplus {\mathbf 27} \oplus {\mathbf
64}~,
\label{twotenrdn}
\eeq
where the singlet ${\mathbf 1}$ is presented, then
\beq
\Hz = \gz \cdot B_{ijk}B^{ijk}.
\eeq
Here Einstein summation convention is adopted, that is the repeated suffix, once as a subscript and once as a superscript, implies the summation.

Next, according to group theory, in the decomposition of ${\mathbf 8} \otimes {\mathbf 8}$, ${\mathbf 8} \otimes {\mathbf 27}$, and ${\mathbf 8} \otimes {\mathbf 64}$, the singlet only exits in that of ${\mathbf 8} \otimes {\mathbf 8}$. In addition, two kinds of breaking effect are to be considered, then the final effective interaction Hamiltonian reads
\beq
\Heff = \gz \cdot B_{ijk}B^{ijk}+ g_m \cdot \Hmass + g_e \cdot \Hchrg~,
\label{effhmtdkp}
\eeq
where
\beq
\Hmass =B_{3jk}B^{3jk} -\frac{1}{3} (B_{ijk}B^{ijk})~,
\label{effhmt11}
\eeq
and
\beq
\Hchrg =B_{1jk}B^{1jk} -\frac{1}{3} (B_{ijk}B^{ijk})~.
\label{effhmt33}
\eeq

Substituting the components of Eqs.~\eref{dkpbyn} and \eref{dkpabyn} into the effective Hamiltonian of Eq.~\eref{effhmtdkp}, we acquire the parametrization for decuplet-decuplet baryon final state as listed in Table~\ref{decupletbynform}.

\begin{table}[hbt]
\caption{\label{decupletbynform}Amplitude parametrization form for decay of $\psp$ or $\jpsi$ into a pair of decuplet baryon, in terms of singlet $A$ (by definition $A=\gz$), as well as the charge-breaking term $D$ (by definition $D=g_m/3$) and the mass-breaking term $\Dp$ (by definition $\Dp=g_e/3$).} \center
\begin{tabular}{ll}\hline \hline
  Final state    & Amplitude form  \\ \hline
  $\DDltpp$      & $A+2D-\Dp$    \\
  $\DDltp$       & $A+D-\Dp$     \\
  $\DDltz$       & $A~~~~~~-\Dp$    \\
  $\DDltn$       & $A-D-\Dp$     \\
  $\SSbp$        & $A+D$       \\
  $\SSbz$        & $A$         \\
  $\SSbm$        & $A-D$       \\
  $\XXbz$        & $A~~~~~~+\Dp$     \\
  $\XXbm$        & $A-D+\Dp$    \\
  $\OOb$         & $A-D+2\Dp$         \\
\hline \hline
\end{tabular}
\end{table}

\begin{table}[hbt]
\caption{\label{octetbynform}Amplitude parametrization form for decay of $\psp$ or $\jpsi$
into a pair of octet baryon, in terms of singlet $A$, as well as symmetric and antisymmetric charge-breaking ($D,F$) and mass-breaking terms ($\Dp,\Fp$). Here $A=\gz$, $D=g_e^{\prime}/3$, $F=-g_e$, $\Dp=-g_m^{\prime}/3$, and $\Fp=g_m$, such a choice is due to the consistency with previous study in Ref.~\cite{zmy2015}.
}\center
\begin{tabular}{ll}\hline \hline
  Final state    & Amplitude form  \\ \hline
  $\ppb$         & $A+D+F-\Dp+\Fp$    \\
  $\nnb$         & $A-2D-\Dp+\Fp$     \\
  $\SSbp$        & $A+D+F+2\Dp$       \\
  $\SSbz$        & $A+D+2\Dp$         \\
  $\SSbm$        & $A+D-F+2\Dp$       \\
  $\XXbz$        & $A-2D-\Dp-\Fp$     \\
  $\XXbm$        & $A+D-F-\Dp-\Fp$    \\
  $\LLb$         & $A-D-2\Dp$         \\
  $\SzLb$,$\SbzL$& $\sqrt{3}D$        \\
\hline \hline
\end{tabular}
\end{table}

Now we consider octet-octet final state. The $SU(3)$ octet baryon is convenient to expressed in the matrix notations~\cite{lichtenberg,hGeorgi}
\beq
{\mathbf B}=
\left(\begin{array}{ccc}
\Sigma^0/\sqrt{2}+\Lambda/\sqrt{6} & \Sigma^+   & p    \\
\Sigma^-  & -\Sigma^0/\sqrt{2}+\Lambda/\sqrt{6} & n    \\
\Xi^-     & \Xi^0            & -2\Lambda/\sqrt{6}
\end{array}\right)~~,
\label{oktbyn}
\eeq
and
\beq
\overline{\mathbf B}=
\left(\begin{array}{ccc}
\overline{\Sigma}^0/\sqrt{2}+\overline{\Lambda}/\sqrt{6}
                    & \overline{\Sigma}^+ & \overline{\Xi}^+    \\
\overline{\Sigma}^- & -\overline{\Sigma}^0/\sqrt{2}+\overline{\Lambda}/\sqrt{6}
                                          & \overline{\Xi}^0    \\
\overline{p}        & \overline{n} & -2\overline{\Lambda}/\sqrt{6}
\end{array}\right)~~.
\label{atoktbyn}
\eeq
The corresponding tensor notations are respectively $B^i_j$ and $\overline{B}^i_j$, where the superscript denotes the row index of matrix and the subscript the column index. According to the decomposition
\beq
{\mathbf 8} \otimes {\mathbf 8} = {\mathbf 1} \oplus {\mathbf 8} \oplus {\mathbf 8}
\oplus {\mathbf 10} \oplus {\mathbf 10^*} \oplus {\mathbf 27}~,
\label{twoeightrdn}
\eeq
the singlet exists which leads to a symmetry conserved interaction, that is
\beq
\Hz = \gz \cdot \overline{B}^i_j B^j_i~~.
\eeq
As far as breaking terms are concerned, the octet-octet final state are more complex than that of decuplet-decuplet one. By virtue of Eq.~\eref{twoeightrdn} it is noted that there are two types of octet: an antisymmetric, or $f$-type, and  a symmetric, or $d$-type, defined respectively by
\beq
([\overline{B} B]_f )^i_j = \overline{B}^i_k B^k_j -\overline{B}^k_j B^i_k~~,
\eeq
and
\beq
([\overline{B} B]_d )^i_j = \overline{B}^i_k B^k_j +\overline{B}^k_j B^i_k
-\frac{2}{3} \delta^i_j \cdot \overline{B}^i_j B^j_i~~.
\eeq
Correspondingly, the each breaking term for decuplet now contains two parts for octet. In addition, no singlet exists in the decomposition of ${\mathbf 8} \otimes {\mathbf 10}$, ${\mathbf 8} \otimes {\mathbf 10^*}$, and ${\mathbf 8} \otimes {\mathbf 27}$, therefore, the final effective interaction Hamiltonian reads

\beq
\left.\begin{array}{rl}
\Heff = & \gz \cdot \overline{B}^i_j B^j_i    \\
        & + g_m \cdot ([\overline{B} B]_f )^3_3  + g_m^{\prime} \cdot ([\overline{B} B]_d )^3_3   \\
        & + g_e \cdot ([\overline{B} B]_f )^1_1  + g_e^{\prime} \cdot ([\overline{B} B]_d )^1_1 ~~~.
\end{array}\right.~~
\label{effhmtctp}
\eeq
Then writing the $\Heff$ in particle form, we acquire the parametrization for octet-octet baryon pair final state as summarized in Table~\ref{octetbynform}.

\begin{table}[hbt]
\caption{\label{okadkbynform}Amplitude parametrization form for decay of $\psp$ or $\jpsi$ into decuplet-octet baryons, in terms of the charge-breaking term $D$ ($D=g_e/2\sqrt{3}$) and the mass-breaking term $\Dp$ ($\Dp=g_m/\sqrt{3}$).} \center
\begin{tabular}{ll}\hline \hline
  Final state    & Amplitude form  \\ \hline
$\overline{\Sigma^*}^- \Sigma^+$ / $\Sigma^{*+} \overline{\Sigma}^-$ & $-2D +\Dp$    \\
$\overline{\Sigma^*}^0 \Sigma^0$ / $\Sigma^{*0} \overline{\Sigma}^0$ & $+D ~-\Dp$    \\
$\overline{\Sigma^*}^+ \Sigma^-$ / $\Sigma^{*-} \overline{\Sigma}^+$ & $~~~~~~~~~\Dp$    \\
$\overline{\Xi^*}^0 \Xi^0$ / $\Xi^{*0} \overline{\Xi}^0$             & $-2D +\Dp$    \\
$\overline{\Xi^*}^+ \Xi^-$ / $\Xi^{*-} \overline{\Xi}^+$             & $~~~~~~~~~\Dp$    \\
$\overline{\Delta}^- p$  /  $\Delta^+ \overline{p}$                  & $~2D$    \\
$\overline{\Delta}^0 n$  /  $\Delta^0 \overline{n}$                  & $~2D$    \\
$\overline{\Sigma^*}^0 \Lambda$ / $\Sigma^{*0} \overline{\Lambda}$   & $-\sqrt{3}D$    \\
 \hline \hline
\end{tabular}
\end{table}

Last, we consider decuplet-octet final state. According to the reduction
\beq
{\mathbf 8} \otimes {\mathbf 10^*} = {\mathbf 8} \oplus {\mathbf 10} \oplus {\mathbf 27}
\oplus {\mathbf 35} ~,
\label{teneightrdn}
\eeq
no singlet exits, so there is no symmetry conserved term in the effective interaction Hamiltonian. All terms come from breaking effects. The octet in the left hand of Eq.~\eref{teneightrdn} is constructed as follows
\beq
O^i_j = \epsilon^{imn} B_{lmj} B^l_n~~,
\eeq
or
\beq
O_i^j = \epsilon_{imn} B^{lmj} \overline{B}_l^n~~,
\eeq
where $\epsilon^{imn}$ (or $\epsilon_{imn}$) is totally antisymmetric tensor. No singlet exists in the decomposition of ${\mathbf 8} \otimes {\mathbf 10}$, ${\mathbf 8} \otimes {\mathbf 27}$, and ${\mathbf 8} \otimes {\mathbf 35}$, therefore, only singlet comes from the product of two octets, and the final effective interaction Hamiltonian reads
\beq
\Heff= g_m O^3_3+ g_e O^1_1~~.
\label{effhmtokdk}
\eeq
The parametrization for octet-decuplet baryon final state is presented in Table~\ref{okadkbynform}.

\subsection{Parametrization of mesonic final state}
The philosophy of parametrization for mesonic final state is similar to that for baryonic final state. We will take vector-pseudoscalar (VP) and pseudoscalar-pseudoscalar (PP) meson pair final states as examples to expound the parametrization process.

$SU(3)$ octet vector meson and pseudoscalar meson are respectively expressed in matrix notations~\cite{aHosaka,sColeman}
\beq
{\mathbf V}=
\left(\begin{array}{ccc}
\roz/\sqrt{2}+\omega/\sqrt{6} & \rop         & \kstp    \\
\rom        & -\roz/\sqrt{2}+\omega/\sqrt{6} & \kstz    \\
\kstm       & \kstzb                         &  -2 \omega/\sqrt{6}
\end{array}\right)~~,
\label{oktvtmsn}
\eeq
and
\beq
{\mathbf P}=
\left(\begin{array}{ccc}
\piz/\sqrt{2}+\eta/\sqrt{6} & \pip         & \kap    \\
\pim        & -\piz/\sqrt{2}+\eta/\sqrt{6} & \kaz    \\
\kam        & \kazb                        & -2 \eta/\sqrt{6}
\end{array}\right)~~.
\label{oktpsmsn}
\eeq
The corresponding tensor notations are respectively $V^i_j$ or $P^i_j$, where the superscript denotes the row index of matrix and the subscript the column index.

\begin{table}[hbt]
\caption{\label{vpmsnform}Amplitude parametrization form for decay of $\psp$ or $\jpsi$ into $V~P$ final state, in terms of singlet $A$ (by definition $A=g_0$), as well as charge-breaking ($D=g_e/3$) and mass-breaking terms ($\Dp=g_m/3$).  } \center
\begin{tabular}{ll}\hline \hline
  Final state    & Amplitude parametrization form  \\ \hline
  $\rho^{\pm}\pi^{\mp}$, $\roz \piz$  & $A+D-2\Dp$    \\
  $K^{*\pm} K^{\mp}$                  & $A+D+\Dp$     \\
  $\kstz\kazb$, $\kstzb \kaz$         & $A-2D+\Dp$    \\
  $\omega \eta$                       & $A-D+2\Dp$    \\
  $\omega \piz$                       & $~~\sqrt{3}D~~~$     \\
  $\roz \eta$                         & $~~\sqrt{3}D~~~$    \\
\hline \hline
\end{tabular}
\end{table}

\begin{table}[hbt]
\caption{\label{ppmsnform}Amplitude parametrization form for decay of $\psp$ or $\jpsi$ into PP final state, in terms of the charge-breaking one ($D=2 g_e$) and the mass-breaking one ($\Dp=-2 g_m$).}
\center
\begin{tabular}{ll}\hline \hline
  Final state    & Amplitude parametrization form  \\ \hline
  $\pip\pim$     & $D$    \\
  $\kap\kam$     & $D+\Dp$     \\
  $\kaz\kazb$    & $~~~~~~\Dp$       \\
\hline \hline
\end{tabular}
\end{table}

For mesonic final state, only octet need to be considered, so similar to the reason for octet baryon pair, both symmetry conserved and symmetry breaking terms come into effective interaction Hamiltonian. Nevertheless, unlike baryon multiplets, meson octet is self-conjugate representation, and particle and anti-particle exist in the same octet, as displayed in Eqs.~\eref{oktvtmsn} and \eref{oktpsmsn}. The charge conjugate symmetry imposes more constraints on the interaction term in Hamiltonian. The analysis indicates that for VP mode only $d$-type octet is permitted while for PP mode only $f$-type octet  permitted~\cite{Haber}. Therefore, the effective Hamiltonian for VP and PP modes are respectively
\beq
\Heff^{VP} = \gz \cdot V^i_j P^j_i
+ g_m \cdot ( [V P]_d )^3_3  + g_e \cdot ([V P]_d )^1_1 ~,
\label{effhmtvpmsn}
\eeq
and
\beq
\Heff^{PP} =
 g_m \cdot ([P \bspaf P]_f )^3_3  + g_e \cdot ([P \bspaf P]_f )^1_1 ~,
\label{effhmtvpmsn}
\eeq
where $a \bspaf b \equiv a (\partial b) - (\partial a) b$, so $a \bspaf b = -b \bspaf a$ and
$a \bspaf a =0$.

With the components given in Eqs.~\eref{oktvtmsn} and \eref{oktpsmsn}, the corresponding parametrization can be obtained and summarized respectively in Table~\ref{vpmsnform} and \ref{ppmsnform}, where the partial derivative symbol $\bspaf$ is suppressed.

A remark is in order here. As can be seen from Eqs.~\eref{oktvtmsn} and \eref{oktpsmsn}, $V$ and $P$ mesons are treated as pure octet particles. However, it is well known that the actual particles are mixing of octet and singlet particles. More pragmatic treatment should combine both octet and singlet representations. One approach is to introduce a synthetical nonet as expounded in Ref.~\cite{Haber}. To integrate such a form into the present formalism, more punctilious and comprehensive work is needed, which will be the content of paper in the future.

\subsection{Comment}
In previous section, all focus are on $\jpsi$ and $\psp$ decays. As a matter of fact, other charmonium singlet, such as $\eta_c$, can also be analyzed similarly. Moreover, the parametrization of other decay modes such as vector-vector meson pair, vector-tensor meson pair can be obtained easily by appropriate change in particle labeling.

The aforementioned general analysis principle can extend to the three body decay as well. However, such an extension is no so appealing, because firstly, more parameters will be introduced, only for certain special case, the concise result is available; secondly, most three-meson final states are dominated by quasi-two-body intermediate states, the non-resonant three body are quite meager; thirdly, many interesting dynamics studies, such as the measurement of the relative phase, can be performed more easily and accurately with two body decays.

In principle, a totally general algebraic analysis of breaking term can be extended for the second-order effect by inserting the appropriate spurion fields in an $SU(3)$-invariant way into the interaction Hamiltonian. Anyway, further analysis will more or less involve the decay dynamics, and therefore more model-dependent.

\section{Experimental section}\label{xct_expmsm}
Since the upgraded Beijing Electron-Positron Collider (BEPCII) and spectrometer detector (BESIII) started data taking in 2008~\cite{bes,yellow}, the largest charmonium data samples in the world were collected, especially the data at $\jpsi$ and $\psp$ resonance peaks, which provide an unprecedented opportunity to acquire useful information for understanding the interaction dynamics of chamonium decay.

However, when analyzing the data taken in $\EE$ collider, the important experimental effects such as the initial state radiative (ISR) correction and the effect due to energy spread of accelerator must be deal with carefully.

\subsection{Born section}

For $\EE$ colliding experiments, there is the inevitable continuum amplitude~\cite{wangp03hepnp}
$$ 
\EE \rightarrow \gamma^* \rightarrow hadrons
$$ 
which may produce the same final state as the resonance decays do. The total Born cross section is therefore reads~~\cite{rudaz,wymcgam,Wang:2005sk}
\beq
\sigma_{B}(s) =\frac{4\pi \alpha^2}{3s}
   |\ag(s)+\aga(s)+\ac(s)|^2~{\cal P}(s)~,
\label{bornxc}
\eeq
which consists of three kinds of amplitudes correspond to (a) the strong interaction ($\ag(s)$) presumably through three-gluon annihilation, (b) the electromagnetic interaction ($\aga(s)$) through the annihilation of $c\overline{c}$ pair into a virtual photon, and (c) the electromagnetic interaction ($\ac(s)$) due to
one-photon continuum process. The phase space factor ${\cal P}$
is expressed as
\beq
{\cal P} = v (3-v^2)/2~,
~~v\equiv \sqrt{1-\frac{(m_{B_1}+m_{\bar{B}_2})^2}{s}}~,
\eeq
where $m_{B_1}$ and $m_{\bar{B}_2}$ are the masses of the baryon and anti-baryon in the final states, and $v$ velocity of baryon in the center-mass-system.

For baryon pair final state, the amplitudes have forms~:
\beq
\ac(s)=\frac{Y}{s}~,
\label{ampac}
\eeq
\beq
\aga(s)=\frac{3Y\Gamma_{ee}/(\alpha\sqrt{s})}
{s-M^2+iM\Gamma_t}~,
\label{ampap}
\eeq
\beq
\ag(s)=\frac{3X\Gamma_{ee}/(\alpha\sqrt{s})}
{s-M^2+iM\Gamma_t}~,
\label{ampag}
\eeq
where $\sqrt{s}$ is the center of mass energy, $\alpha$ the QED fine structure constant; $M$ and $\Gamma_t$ are the mass and the total width of $\psp$ or $\jpsi$; $\Gamma_{ee}$ is the partial width to $\EE$.
$X$ and $Y$ are functions of amplitude parameters $A,D,F,\Dp$, and $\Fp$ listed in Table~\ref{octetbynform}, viz.
\beq
Y=Y(D,F)~,
\label{defy}
\eeq
\beq
X=X(A,\Dp,\Fp) e^{i\phi}~.
\label{defx}
\eeq
The special form of $X$ or $Y$ depends on decay mode, as examples, for $\ppb$ decay mode, $X=A-\Dp+\Fp$ and $Y=D+F$ while for $\XXbm$ decay mode, $X=A-\Dp-\Fp$ and $Y=D-F$, according to the parametrization forms in Table~\ref{octetbynform}. In principle, the parameters listed in Table~\ref{octetbynform} could be complex arguments, each with a magnitude together with a phase, so there are totally ten parameters which are too many for nine octet-baryon decay modes. To make the following analysis practical, it is assumed that there is not relative phases among the strong-originated amplitudes $A$, $\Dp$, $\Fp$, and electromagnet amplitudes $D$, $F$; the sole phase (denoted by $\phi$ in Eq.~\eref{defx} ) is between the strong and electromagnet interactions, that is between $X$ and $Y$, as indicated in Eqs.~\eref{defx} and \eref{defy}, where $A$, $D$, $F$, $\Dp$, and $\Fp$ are treated actually as real variables.

\subsection{Observed section}

In $\EE$ collision, the Born order cross section is
modified by the initial state radiation in the way~\cite{rad.1}
\begin{equation}
\sigma_{r.c.} (s)=\int \limits_{0}^{x_m} dx
F(x,s) \frac{\sigma_{Born}(s(1-x))}{|1-\Pi (s(1-x))|^2}~,
\label{eq_isr}
\end{equation}
where $x_m=1-s'/s$. $F(x,s)$ is the radiative function been calculated to an accuracy of
0.1\%~\cite{rad.1,rad.2,rad.3}, and $\Pi(s)$ is the vacuum polarization factor. In the
upper limit of the integration, $\sqrt{s'}$ is the experimentally required minimum
invariant mass of final particles.
If $x_m=1$, it corresponds to no requirement for invariant mass; if $x_m=0.2$, it corresponds to invariant mass cut of 3.3~GeV for $\psp$ resonance. The concrete value of $x_m$ should be determined by the cut of invariant mass, which is adopted in actual event selection.


The $\EE$ collider has a finite energy resolution which is much wider than the intrinsic width of narrow
resonances such as $\psp$ and $\jpsi$~\cite{besscan95,besscan02}. Such an energy resolution is usually a Gaussian distribution~\cite{awChao}:
$$
G(W,W^{\prime})=\frac{1}{\sqrt{2 \pi} \Delta}
             e^{ -\frac{(W-W^{\prime})^2}{2 {\Delta}^2} },
$$
where $W=\sqrt{s}$ and $\Delta$, a function of energy, is the standard deviation of Gaussian distribution. The experimentally observed cross section is the radiative corrected cross section folded with the energy resolution function
\begin{equation}
\sigma_{obs} (W)=\int \limits_{0}^{\infty}
        dW^{\prime} \sigma_{r.c.} (W^{\prime}) G(W^{\prime},W)~.
\label{eq_engsprd}
\end{equation}

\begin{table*}[bth]
\caption{\label{tab_expcdn}Breakdown of experiment conditions correspond to different detectors and accelerators. The energy spread is the effective one, according to which the calculated maximum cross section satisfies the relation $N_{tot} =\sigma_{max} \cdot {\cal L}$. The number with star ($\ast$) is the equivalent luminosity calculated by relation ${\cal L}=N_{tot}/\sigma_{max}$. }
\begin{ruledtabular}
\begin{tabular}{llllllll}
         &             &C.M. Energy &Data Taking & Maximum   & Total & Integral   & \\
Detector & Accelerator &  Spread    &Position\footnote{
\begin{minipage}{13cm}\mbox{}The data taking position is the energy which
yield the maximum inclusive hadronic cross section. \end{minipage} }
                                                 & section   & event & luminosity &Ref. \\
         &             & (MeV)      & (GeV)      &  (nb)     &($\times 10^6$)
                                                                     & (pb$^{-1}$) & \\ \hline
CLEO-c  & CESR    &   1.68  & 3.68627   & 557.23   & 3.08    & 5.63       & \cite{zjycleo05} \\
        &         &   1.68  & 3.67      & $-$      & $- $    & 20.70      & \\
CLEO-c  & CESR    &   1.821 & 3.68629   & 510.54   & 24.5    & 48         & \cite{zjycleoc17} \\
BES     & BEPC    &   1.23  & 3.68623   & 712.9    & 3.95    & 5.541$\ast$& \cite{Bai:2000ye}\\
BES II  & BEPC    &   1.23  & 3.68623   & 712.9    & 14.0    & 19.72      & \cite{tnpsp2004}\\
        &         &   1.23  & 3.65      & $-$      & $- $    & 6.42       & \cite{lum2004}\\
BESIII  & BECPII  &   1.343 & 3.68624   & 662.16   & 107.0   & 161.63     & \cite{tnpsp2018}\\
        &         &   1.343 & 3.65      & $-$      & $- $    & 43.88      & \cite{tnpsp2018} \\
        &         &   1.318 & 3.68624   & 672.74   & 341.1   & 506.92     & \cite{tnpsp2018} \\
        &         &   1.324 & 3.68624   & 670.17   & 448.1   & 668.55     & \cite{tnpsp2018}\\
BESIII   & BECPII    & 1.131   & 3.097014   &  2808.63  & 223.7  & 79.63  & \cite{tnjps2012}\\
            &        & 1.131   & 3.08       &  $-$      & $-$    & 30.84  & \cite{tnjps2017} \\
            &        & 0.898   & 3.096990   &  3447.87  & 1086.9 & 315.02 & \cite{tnjps2017} \\
            &        & 0.937   & 3.096993   &  3320.35  & 1310.6 & 394.65  & \cite{tnjps2017} \\
BES II   & BEPC      & 0.85    & 3.09700    & 3631.8    & 57.7  & 15.89$\ast$ & \cite{tnjps2003}\\
MARK II  & SPEAR     & 2.40    & 3.097108   & 1429.3    & 1.32  & 0.924$\ast$ & \cite{mrk2bbdk}\\
MARK III & SPEAR     & 2.20    & 3.097121   & 1541.6    & 2.71  & 1.758$\ast$ & \cite{pdg1996,mrk3pp85}\\
 DM  II  & DCI       & 1.98    & 3.097114   & 1702.0    & 8.6   & 5.053$\ast$ & \cite{dm2bbdka}\\
\end{tabular}
\end{ruledtabular}
\end{table*}

In fact, as pointed out in Ref.~\cite{wymcgam}, the radiative correction and the energy spread of the collider are two important factors, both of which reduce the height of the resonance and shift the position of the maximum cross section. Although the ISR are the same for all $\EE$ experiments, the energy spread is quite different for different accelerators, even different for the same accelerator at different running periods. As an example, for the CLEO data used in this paper, the energy spread varies due to different accelerator lattices~\cite{YELLOWBOOK}: one (for CLEO III detector) with a single wiggler
magnet and a center-of-mass energy spread $\Delta$=1.5~MeV, the other (for CLEOc detector) with the first half of its full complement (12) of wiggler magnets and $\Delta$=2.3~MeV. The two $\Delta$'s lead to two maximum total cross sections 602 nb and 416 nb, respectively. All these subtle effects must be taken into account in data analysis. In the following analysis all data are assumed to be taken at the energy point which yields the maximum inclusive hadron cross sections in stead of the nominal resonance mass~\cite{wymcgam,wymhepnp}. Besides the factors considered above, the resonance parameters can also affect the evaluation results. Since the present central values of resonance parameters can be obviously distinct from those of some time before, the calculated maximum inclusive hadron cross sections will consequently different. In order to ensure the relation $N_{tot} =\sigma_{max} \cdot {\cal L}$, some adjustments are needed. The principle is as follows: if the luminosity is available, the energy spread will be tuned to give consistent maximum cross section; otherwise, the effective luminosity is evaluated by the relation ${\cal L}=N_{tot}/\sigma_{max}$ by virtue of the corresponding accelerator parameters. All experimental details are summarized in Table~\ref{tab_expcdn}, which are crucial for the following data analysis. At last, the resonance parameters adopted in this monograph for $\jpsi$ and $\psp$ are respectively~\cite{pdg2020}
\beq
\begin{array}{rcl}
   M_R &=&3096.900\pm 0.006 ~~\mbox{MeV }, \\
   \Gamma_t &=&92.9\pm 2.8~~\mbox{keV }, \\
   \Gamma_{ee}&=& 5.53\pm 0.10~~\mbox{keV };
\end{array}
\label{jpsirnsprt}
\eeq
and
\beq
\begin{array}{rcl}
   M_R &=&3686.10\pm 0.06 ~~\mbox{MeV }, \\
   \Gamma_t &=&294\pm 8~~\mbox{keV }, \\
   \Gamma_{ee}&=& 2.33\pm 0.04~~\mbox{keV }.
\end{array}
\label{psiprnsprt}
\eeq

\section{Data analysis}\label{xct_fsfit}
There are great many experimental results of $\psp$ and $\jpsi$ decay to octet baryon pair, decuplet baryon pair, decuplet-octet baryon final state, and pseudoscalar-pseudoscalar meson final state. The plenty of data information lays the foundation for the systematically analysis in the light of our parametrization scheme.

Since our analysis involves the experimental details as indicated by description in the preceding section, some measurements are not suitable in the following study due to the lack of necessary information of detectors and/or accelerators. In addition, at different energy point, the status parameters of accelerators are also distinctive, so the studies of phase angle for $\psp$ and $\jpsi$ decay are performed separately for the sake of clarity.

\begin{table*}[bth]
\caption{\label{tab_pspdt_oct}Experimental data of $\psp$ decaying to octet baryon pair final states. For branching fractions, the first uncertainties are statistical, and the second are systematic. For the other quantities, the errors are merely statistical.}
\begin{ruledtabular}
\begin{tabular}{lllll}
  Mode  & $N^{obs}$    &  Efficiency  &Branching ratio  &Detector \\
        &  (peak)      &  (\%)        & ($\times 10^{-4}$)    & \\ \hline
$\ppb$  &$556.5\pm 23.3$ &$66.6\pm 2.8$ &$2.87\pm 0.12\pm 0.15$ & CLEO~\cite{zjycleo05} \\
        &$1618.2\pm 43.4$&$34.4\pm 0.2$ &$3.36\pm 0.09\pm 0.25$ & BESII~\cite{zjybes2} \\
        &$18984\pm 138 $ &$58.1\pm 0.4$ &$3.05\pm 0.02\pm 0.12$ & BESIII~\cite{zjybes3} \\
        &$4475\pm 78  $  &$63.1\pm 1.0$ &$3.08\pm 0.05\pm 0.18$ & CLEO-c~\cite{zjycleoc17} \\

$\nnb$  &$6056  \pm 117$ &$18.5\pm 0.4$ &$3.06\pm 0.06\pm 0.14$ & BESIII~\cite{zjybes3} \\

$\SSbp$ &$34.2 \pm 5.86$ &$4.1\pm 0.8$  &$2.57\pm 0.44\pm 0.68$ & CLEO~\cite{zjycleo05} \\
        &$1874 \pm 46  $ &$33.0\pm 0.9$ &$2.31\pm0.06\pm 0.10$ & CLEO-c~\cite{zjycleoc17} \\
        &$5447 \pm 76  $ &$4.83\pm 0.08$ &$2.52\pm0.04\pm 0.09$ & BESIII~\cite{bes21spsm} \\

$\SSbz$ &$58.0 \pm 7.7$  &$7.2\pm 1.0$  &$2.63\pm 0.35\pm 0.21$ & CLEO~\cite{zjycleo05}\\
        &$59.1 \pm 9.1$  &$1.80\pm0.05$ &$2.35\pm0.36\pm 0.32$ & BESII~\cite{zjybes2} \\
        &$6612\pm 82  $  &$6.04\pm 0.08$&$2.44\pm0.03\pm 0.11$ & BESIII~\cite{zjybes3lmd} \\
        &$2645\pm 56  $  &$48.6\pm 1.1$ &$2.22\pm0.05\pm 0.11$ & CLEO-c~\cite{zjycleoc17} \\

$\XXbz$ &$19.0\pm 4.4 $  &$2.4\pm0.6$ &$2.75\pm0.64\pm0.61 $ & CLEO~\cite{zjycleo05} \\
        &$10839\pm 123 $ &$8.86\pm 0.07$&$2.73\pm 0.03\pm 0.13$ & BESIII~\cite{zjybes3xxbz} \\
        &$1242 \pm 38 $  &$25.6\pm0.8$  &$1.97\pm0.06\pm0.11 $ & CLEO-c~\cite{zjycleoc17} \\
$\XXbm$ &$63.0\pm 8.0 $  &$8.6\pm1.1$ &$2.38\pm0.30\pm0.21 $ & CLEO~\cite{zjycleo05}\\
        &$67.4\pm 8.9 $  &$1.59\pm0.04$ &$3.03\pm0.40\pm 0.32$ & BESII~\cite{zjybes2} \\
        &$3580 \pm 61 $  &$48.2\pm 0.8$ &$3.03\pm0.05\pm0.14 $ & CLEO-c~\cite{zjycleoc17} \\
        &$5336.7\pm82.6$ &$18.04\pm0.04$&$2.78\pm0.05\pm0.14 $ & BESIII~\cite{zjybes3xxb} \\
$\LLb$  &$203.5\pm 14.3$ &$20.1\pm1.4$&$3.28\pm0.23\pm0.25 $ & CLEO~\cite{zjycleo05} \\
        &$337.2\pm 19.9$ &$7.10\pm0.11$ &$3.39\pm0.20\pm0.32 $ & BESII~\cite{zjybes2} \\
        &$31119\pm 187$  &$17.49\pm0.24$&$3.97\pm0.02\pm0.12 $ & BESIII~\cite{zjybes3lmd} \\
        &$6531 \pm 82 $  &$71.6\pm1.0$  &$3.71\pm0.05\pm0.15 $ & CLEO-c~\cite{zjycleoc17} \\
$\SzLb+\SbzL$
        &$30  \pm 5   $  &$9.9\pm 1.9$ &$0.123\pm 0.023\pm0.008$ & CLEO-c~\cite{zjycleoc17} \\
        &$63.8 \pm 10.2$ &$7.4\pm 1.3$ &$0.0160\pm 0.0031\pm0.0013$ & BESIII~\cite{bes21xlacc} \\
\end{tabular}
\end{ruledtabular}
\end{table*}

\begin{table*}[bth]
\caption{\label{tab_jpsidtzjy}Experimental data of $\jpsi$ decaying to octet baryon pair final states. For branching fractions, the first uncertainties are statistical, and the second are systematic. For the other quantities, the errors are merely statistical. }
\begin{ruledtabular}
\begin{tabular}{lllll}
  Mode  & $N^{obs}$       & Efficiency  & Branching Ratio   & Detector     \\
        &  (peak)         & (\%)      &     ($\times 10^{-4}$)    &     \\ \hline
$\ppb$  &$63316 \pm  281$ &$48.53\pm 0.31$&$22.6 \pm 0.1 \pm 1.4 $ & BESII~\cite{besbbdka} \\
        &$1420 \pm  46$   &$49.7\pm 1.6$  &$21.6 \pm 0.7 \pm 1.5 $ & MARKII~\cite{mrk2bbdk} \\
        &$314651 \pm 561$ &$66.1\pm 0.17$ &$21.12\pm 0.04\pm 0.31$ & BESIII~\cite{Ablikim:2012eu} \\
$\nnb$  &$35891 \pm 211$  &$7.69\pm 0.06$ &$20.7 \pm0.1  \pm 1.7 $ & BESIII~\cite{Ablikim:2012eu} \\
$\SSbp$ &$399   \pm 26 $  &$0.45\pm 0.03$ &$15.0 \pm1.0  \pm 2.2 $ & BESII~\cite{besbbdkc} \\
        &$86976 \pm 314 $ &$6.26\pm 0.03$ &$10.61\pm0.04\pm 0.36$ & BESIII~\cite{bes21spsm} \\

$\SSbz$ &$1779 \pm 54$    &$2.31\pm 0.07$ &$13.3 \pm0.4  \pm 1.1 $ & BESII~\cite{besbbdkb} \\
        &$90 \pm  10$     &$4.3 \pm 0.4$  &$15.8 \pm1.6  \pm 2.5 $ & MARKII~\cite{mrk2bbdk} \\
        &$884 \pm 30$     &$9.70\pm 0.37$ &$10.6 \pm0.4  \pm2.3  $ & DMII~\cite{dm2bbdka}   \\
        &$111026 \pm335 $ &$7.28\pm 0.08$ &$11.64\pm0.04 \pm0.23 $ & BESIII~\cite{zjybes3lmd} \\
$\XXbz$ &$206   \pm 20 $  &$0.29\pm 0.03$ &$12.0 \pm1.2  \pm 2.1 $ & BESII~\cite{besbbdkc} \\
        &$134846 \pm 437$ &$8.83\pm 0.07$ &$11.65\pm 0.04\pm 0.43$ & BESIII~\cite{zjybes3xxbz} \\
$\XXbm$ &$194 \pm  14$    &$12.9\pm 0.9$  &$11.4 \pm 0.8 \pm2.0  $ & MARKII~\cite{mrk2bbdk} \\
        &$132 \pm  12$    &$2.20\pm 0.19$ &$7.0 \pm 0.6  \pm 1.2 $ & DMII~\cite{dm2bbdkb}   \\
        &$961 \pm  35$    &$1.83\pm 0.03$ &$9.0 \pm 0.3 \pm 1.8$   & BESII~\cite{bes2012llpp} \\
        &$42810.7\pm231.0$&$18.40\pm 0.04$&$10.40\pm 0.06\pm 0.74$ & BESIII~\cite{zjybes3xxb} \\
$\LLb$  &$8887 \pm  132$  &$7.55\pm 0.11$ &$20.3 \pm0.3  \pm1.5  $ & BESII~\cite{besbbdkb} \\
        &$365  \pm  19$   &$17.6\pm 0.9$  &$15.8 \pm0.8  \pm 1.9 $ & MARKII~\cite{mrk2bbdk} \\
        &$1847 \pm  67$   &$15.6\pm 0.57$ &$13.8 \pm0.5  \pm 2.0 $ & DMII~\cite{dm2bbdka}   \\
        &$440675\pm 670 $ &$17.30\pm 0.20$&$19.43\pm0.03 \pm0.33 $ & BESIII~\cite{zjybes3lmd} \\
$\SzLb$ &$305 \pm 24$     &$8.86\pm 0.67$ &$0.146\pm0.011\pm0.012$ & BESIII~\cite{Ablikim:2012bw}   \\
$\SbzL$ &$234 \pm 20$     &$7.19\pm 0.54$ &$0.137\pm0.012\pm0.011$ & BESIII~\cite{Ablikim:2012bw}   \\
\end{tabular}
\end{ruledtabular}
\end{table*}

\subsection{Octet-Octet mode}
The earlier experimental measurements concerned with $\psp$ decaying to octet-baryon pair final state are
contained in Refs.~\cite{Feldman:1977nj,Brandelik:1979hy,zjymk1,zjyfeni,Bai:2000ye}. The results of
Refs.~\cite{Feldman:1977nj} and \cite{Brandelik:1979hy} are presented three decades ago and moreover only one branching fraction (for $\ppb$) and two upper limits (for $\LLb$ and $\XXbm$) are given. In Ref.~\cite{Bai:2000ye}, the branching fractions of $\ppb$, $\LLb$, $\SSbz$ and $\XXbm$  are obtained based on 4 million $\psp$ events. Besides larger uncertainties, the central values are also rather distinctive from the latter more accurate measurements. Therefore, we will focus on the measurement results after 2001.

For recent results, Refs.~\cite{dobbs14} and~\cite{zjycleoc17} have the similar analysis, but in the latter paper, the efficiency of hyperon identification have substantially improved (by factors 3-5). For this reason, only the latter results are adopted for the analysis herein. For $\SzLb+\SbzL$ final state, there are two measurements from Refs.~\cite{zjycleoc17} and~\cite{bes21xlacc}. However, the branching fraction of CLEO is almost one order of magnitude larger then that of BESIII. Here we follow the strategy of PDG2022~\cite{pdg2022}, only adopt the result of BESIII.

Reference~\cite{aubert07} provides the branching fractions of $\LLb$ and $\SSbz$ for $\jpsi$ and $\psp$ decays, but the results are obtained by using the initial state radiation technique, which is too different to be merged with other information. These data are not utilized in the following analysis.
All measurement results that are to be utilized are summarized in Table~\ref{tab_pspdt_oct}.

As far as the aforementioned principle is concerned, the energy spread will be tuned to give the maximum cross section that can satisfy the relation $N_{tot} =\sigma_{max} \cdot {\cal L}$. CLEO data~\cite{zjycleo05} are composed of two sets, one with luminosity 2.74 pb$^{-1}$ and the other 2.89 pb$^{-1}$, which are taken with energy spreads 1.5~MeV and 2.3~MeV, respectively. In the following analysis, the data is treated as one set with total luminosity 5.63 pb$^{-1}$ corresponding to the effective energy spread 1.68~MeV as displayed in Table~\ref{tab_pspdt_oct}. It it worthy of noticing that unlike branching fraction evaluation, the contribution due to QED continuum should not be subtracted from the observed number of events, since the QED contribution is included in the observed cross section calculation.

Chi-square method is adopted to fit the experiment data. The estimator is constructed as
\beq
\chi^2= \sum\limits_i
\frac{[N_i - n_i(\vec{\eta})]^2}{(\delta N_i)^2}~,
\label{chisqbb}
\eeq
where $N$ with the corresponding error ($\delta N$) denotes the experimentally measured number of events while $n$ the theoretically calculated number of events~:
\beq
n={\cal L} \cdot \sigma_{obs} \cdot \epsilon~,
\label{eq_defsig}
\eeq
where ${\cal L}$ is the integrated luminosity and $\sigma_{obs}$ the observed cross section calculated according to formula~\eref{eq_engsprd}, which contains the parameters to be fit, such as $A$, $D$, $F$, $\Dp$, $\Fp$, and the phase angle $\phi$. All these parameters are denoted by the parameter vector $\vec{\eta}$ in Eq.~\eref{chisqbb}. $\epsilon$ is the synthetic efficiency that can be expressed as $\epsilon_{MC} \cdot \Pi {\cal B}_{i}$, where $\epsilon_{MC}$ is the efficiency due to Monte Carlo simulation and $\Pi {\cal B}_{i}$ the product of the branching fractions of all intermediate states.

All observed numbers of events together with the corresponding efficiencies displayed in Table~\ref{tab_pspdt_oct} are employed as input information, the fitting results are given as follows:
\beq
\begin{array}{rcl}
   \phi  &=&-94.59^\circ\pm 1.31^\circ,\mbox{ or } ~+85.42^\circ\pm 2.25^\circ; \\
       A &=&~~2.887 \pm 0.008~, \\
     \Dp &=& -0.157 \pm 0.003~, \\
     \Fp &=&~~0.199 \pm 0.013~, \\
      D  &=&~~0.057 \pm 0.002~, \\
      F  &=& -0.467 \pm 0.049~, \\
  f_{bes2}&=&~~0.848 \pm 0.022~, \\
  f_{bes3a}&=&~~0.766 \pm 0.010~, \\
  f_{cleoa}&=&~~0.723 \pm 0.025~, \\
  f_{cleob}&=&~~0.840 \pm 0.009~.
\end{array}
\label{fitpspoo}
\eeq
The scan for each parameter discloses two minima of $\phi$ with opposite sign, while all other parameters remain the same.

All data can be grouped into five sets: two from BESIII, one with total luminosity 668.55 pb$^{-1}$, the other with luminosity 161.63 pb$^{-1}$; two from CLEO, one with total luminosity 48 pb$^{-1}$, the other with luminosity 5.63 pb$^{-1}$; one from BESII with luminosity 19.72 pb$^{-1}$. There might be some systematic difference among those data sets, so normalization factors are introduced to take into account unclear effects. However, only four relative (relative to the greatest data set of BESIII) factors of luminosity are introduced with the belief that the relative relations of measurements of each experiment group is more reliable than the corresponding absolute values. The fit values of four factors $f_{bes2}$, $f_{bes3a}$, $f_{cleoa}$, and $f_{cleob}$ indicate that there indeed exists certain obvious difference, since the inconsistencies of these experiments from the highest precision one are at the level of 20\%.

Now we turn to the analysis of $\jpsi$ decay.

There are lots of measurements for octet baryon pair decays at $\jpsi$ region. However, many of measurements have been performed almost twenty or even more than forty years ago~\cite{Peruzzi:1977pb}-\cite{Bai:1998fu}. The recent experiment results are mainly from BES~\cite{besbbdka,besbbdkb,besbbdkc} and BESIII~\cite{Ablikim:2012eu,Ablikim:2012bw} collaborations. Besides these data, the data from MARKII~\cite{mrk2bbdk} and DMII~\cite{dm2bbdka,dm2bbdkb} are adopted, since the numbers of events from these two experiment group are considerable large and
the more information of distinctive decay modes are also provided.
However, the recent results from Belle~\cite{Wu06bell} are not adopted since the branching fractions of $\jpsi \to \ppb$ and $\LLb$ are measured from B meson decay, whose feature is too different to be merged with other information.
All data used in this analysis are summarized in Table~\ref{tab_jpsidtzjy}.

The minimization estimator for $\jpsi$ is similar to that of $\psp$ as defined in Eq.~\eref{chisqbb}. Anyway, for $\jpsi$ data there is lack of the details about some detectors, especially those of luminosity. Therefore, it is difficult to deal with all data consistently and accurately. To alleviate
the possible inconsistence among the data from different experiment group, four relative (relative to the quantity of BES) normalized factors of luminosity
are introduced.

The fitting results of parameters are listed as follows:
\beq
\begin{array}{rcl}
  \phi  &=&-84.81^\circ\pm 0.70^\circ,\mbox{ or } ~+95.19^\circ\pm 0.70^\circ; \\
       A &=&~~1.676 \pm 0.004~, \\
     \Dp &=& -0.106 \pm 0.001~, \\
     \Fp &=&~~0.192 \pm 0.002~, \\
      D  &=& -0.102 \pm 0.002~, \\
      F  &=&~~0.092 \pm 0.013~; \\
  f_{mk2}&=&~~0.928 \pm 0.025~, \\
  f_{dm2}&=&~~0.787 \pm 0.022~, \\
  f_{bes3a}&=&~~1.011 \pm 0.005~, \\
  f_{bes3b}&=&~~0.930 \pm 0.004~.
\end{array}
\label{fitjpsioo}
\eeq
Here four factors $f_{mk2}$, $f_{dm2}$, $f_{bes3a}$, and $f_{bes3b}$ are used to normalize
the total integral luminosity for experiments at AMRKII, DMII, and BESIII, respectively. The fit values indicate that the largest inconsistencies of these experiments from that of BES can reach to more than 20\%.

According to the fitting results of octet-octet baryon pair final states, there is a large relative phase around $90^\circ$ between the strong and electromagnetic amplitudes. As far as other parameters are concerned, the relative strengthen of $SU(3)$-conserving effect (denoted by $A$) is almost one order of magnitude greater than that of $SU(3)$-breaking effect (denoted by $D,~F,~\Dp,$ and $\Fp$), just as it is expected. Comparing parameters of $\jpsi$ and $\psp$ decays, the pattern of relative strength is similar but is not exactly the same, which implies some distinctive features of their decay mechanisms.

\begin{table*}[bth]
\caption{\label{tab_jpsapspdt_dec}Experimental data of $\psp$ and $\jpsi$ decaying to decuplet baryon pair final states. For branching fractions (${\cal B}_{\psp}\times 10^{5}/{\cal B}_{\jpsi}\times 10^{3}$), the first uncertainties are statistical, and the second are systematic. For the other quantities, the errors are merely statistical. The efficiency with star ($\ast$) is evaluated by virtue of the observed number of events $N^{obs}$, the total number of resonance events, and the corresponding of branching fraction. }
\begin{ruledtabular}
\begin{tabular}{lllll}
  Mode  & $N^{obs}$       & Efficiency   & Branching Ratio           & Detector     \\
        &  (peak)         & (\%)         &($\times 10^{-5}/10^{-3}$) &     \\ \hline
\multicolumn{5}{c}{$\psp$ decay} \\ \hline
$\DDltpp$ &$ 157 \pm  13$ &$ 31 \pm  2$    &$12.8\pm 1.0\pm 3.4$ & BES~\cite{Bai:2000ye}   \\
$\Sigma(1385)^+ \overline{\Sigma}(1385)^-$
      &$1469.9 \pm 94.6$  &$16.45\pm 0.04$ &$8.4 \pm 0.5\pm 0.5$ & BESIII~\cite{zjybes3xxb} \\
          &$ 14 \pm  4$   &$3.29 \pm 0.20$ &$ 11 \pm  3 \pm  3$  & BES~\cite{Bai:2000ye}  \\
$\Sigma(1385)^0 \overline{\Sigma}(1385)^0$
         &$ 2214 \pm 149$ &$7.21\pm 0.02$  &$6.9 \pm 0.5\pm 0.5$ & BESIII~\cite{zjybes3xxbz} \\
$\Sigma(1385)^- \overline{\Sigma}(1385)^+$
       &$1375.5 \pm 97.8$ &$15.12\pm 0.04$ &$8.5 \pm 0.6\pm 0.6$ & BESIII~\cite{zjybes3xxb} \\
$\Xi(1530)^0 \overline{\Xi}(1530)^0$
  &$ 1475.5 \pm 34.1$ &$4.86^{\ast}\pm 0.10$ &$6.77\pm 0.14\pm 0.39$ & BESIII~\cite{bes3xssaxsx21} \\
$\Xi(1530)^- \overline{\Xi}(1530)^+$
  &$ 2533.5\pm 87.2$ &$4.94^{\ast}\pm 0.17$  &$11.45\pm0.40 \pm0.59$ & BESIII~\cite{zjybes3xxm} \\
$\OOb$  &$ 326 \pm  19$ &$25.8\pm 1.5$  &$5.2 \pm 0.3 \pm0.3 $&  CLEO-c~\cite{zjycleoc17} \\
        &$ 27  \pm  5$  &$2.32\pm 0.44$ &$4.7 \pm 0.9 \pm0.5 $&  CLEO-c~\cite{dobbs14} \\
        &$ 10.8\pm 3.5$ &$1.5 \pm 0.5$  &$4.80 \pm 1.56\pm 1.30$& BESIII~\cite{bes2omg12} \\
        &$ 4035 \pm  76$ &$15.39^{\ast} \pm 0.32$ &$5.85 \pm 0.12\pm 0.25$& BESIII~\cite{zjybes3omg} \\ \hline
\multicolumn{5}{c}{$\jpsi$ decay} \\ \hline
$\DDltpp$ &$ 233 \pm  19$ &$16.0 \pm 1.3$ &$ 1.10\pm  0.09\pm 0.28$ & MRKII~\cite{mrk2bbdk}  \\
$\Sigma(1385)^+ \overline{\Sigma}(1385)^-$
 &$52522.5 \pm 595.9$ &$18.67 \pm 0.04$   &$1.258\pm 0.014\pm 0.078$ & BESIII~\cite{zjybes3xxb} \\
          &$  68 \pm  16$ &$ 5.0 \pm 1.2$ &$1.03 \pm 0.24 \pm 0.25 $ & MRKII~\cite{mrk2bbdk} \\
          &$ 1033\pm  56$ &$1.18 \pm 0.03$&$1.50 \pm 0.08 \pm 0.38 $ & BESII~\cite{bes2012llpp} \\
  &$ 754 \pm  27$ &$ 7.37^{\ast} \pm 0.25$&$1.19 \pm 0.04 \pm 0.25 $ & DMII~\cite{dm2bbdkb} \\
$\Sigma(1385)^0 \overline{\Sigma}(1385)^0$
&$102762 \pm  852$ &$ 7.32\pm 0.02$       &$1.071\pm 0.009\pm 0.082$ & BESIII~\cite{zjybes3xxbz}  \\
$\Sigma(1385)^- \overline{\Sigma}(1385)^+$
&$42594.8 \pm 466.8$ &$17.38 \pm 0.04$    &$1.096\pm 0.012\pm 0.071$ & BESIII~\cite{zjybes3xxb} \\
          &$  56 \pm  14$ &$ 5.0 \pm 1.0$ &$0.86 \pm 0.18 \pm 0.22 $ & MRKII~\cite{mrk2bbdk} \\
          &$ 835 \pm  50$ &$ 1.17\pm 0.02$&$1.23 \pm 0.07 \pm 0.30 $ & BESII~\cite{bes2012llpp} \\
 &$ 631 \pm  25$ &$ 7.34^{\ast} \pm 0.25$ &$1.00 \pm 0.04 \pm 0.21 $ & DMII~\cite{dm2bbdkb} \\
\end{tabular}
\end{ruledtabular}
\end{table*}

\subsection{Decuplet-Decuplet mode}
The experimental data of $\psp$ and $\jpsi$ decay to decuplet baryon pair final states are summarized in Table~\ref{tab_jpsapspdt_dec}. $\Sigma(1385)$ and $\Xi(1530)$ are also denoted by ${\Sigma^*}$ and ${\Xi^*}$ as displayed in Table~\ref{okadkbynform}

For $\psp \to \OOb$ decay, Refs.~\cite{dobbs14} and~\cite{zjycleoc17} have the similar analysis, but in the latter paper, the efficiency of hyperon identification have substantially improved (by factors 3-5), therefore only the latter result is adopted for the analysis. In addition, in Ref.~\cite{zjycleo05} based on 4 million $\psp$ events observed are only 4 events, whose statistic is too low to be adopted.
Moreover, the measurement of Ref.~\cite{bes2omg12} is not adopted either due to the low statistic.
The same criterion is also applicable for BES measurement of $\Sigma(1385)^+ \overline{\Sigma}(1385)^-$ final state~\cite{Bai:2000ye}.

For $\jpsi$ decay to decuplet baryon pair final states, the experimental information is comparative limited, therefore all measurement results are utilized for parameters fitting.

The fitting philosophy of $\psp$ and $\jpsi$ decay to decuplet baryon pair final states is the similar to that of $\psp$ and $\jpsi$ decay to octet baryon pair final states. A few normalization factors are introduced to take into account some unclear systematic difference among experimental data sets.  The fitted parameters for $\psp$ decay are listed as follows:
\beq
\begin{array}{rcl}
  \phi  &=&-75.51^\circ\pm 4.87^\circ,\mbox{ or } ~+104.49^\circ\pm 4.91^\circ; \\
       A &=&~1.621 \pm 0.031~, \\
     \Dp &=&-0.171 \pm 0.031~, \\
      D  &=&~0.593 \pm 0.060~, \\
  f_{cleo}&=&~0.806 \pm 0.051~, \\
  f_{bes}&=&~0.768 \pm 0.146~, \\
  f_{bes3a}&=&~0.702 \pm 0.058~.
\end{array}
\label{fitpspdd}
\eeq

The fitted parameters for $\jpsi$ decay are listed as follows:
\beq
\begin{array}{rcl}
  \phi  &=&-96.28^\circ\pm 17.23^\circ~,\mbox{ or } ~+83.27^\circ\pm 11.38^\circ ; \\
       A &=&~1.789 \pm 0.007~~, \\
     \Dp &=&~0.347 \pm 0.461~~, \\
      D  &=&~0.389 \pm 0.542~~, \\
  f_{mk2}&=&~0.719 \pm 0.151~~, \\
  f_{dm2}&=&~0.840 \pm 0.111~~, \\
  f_{bes2}&=&~1.058 \pm 0.145~~, \\
  f_{bes3a}&=&~0.902 \pm 0.121~~.
\end{array}
\label{fitjpsidd}
\eeq

According to the fitting results of  decuplet final states, there is a large relative phase around $90^\circ$ between the strong and electromagnetic amplitudes. As far as other parameters are concerned, the relative strengthen of $SU(3)$-conserving effect (denoted by $A$) is fairly greater than that of $SU(3)$-breaking effect (denoted by $D$ and $\Dp$), just as it is expected. Comparing parameters of $\jpsi$ and $\psp$ decays, the pattern of relative strength is similar but is not exactly the same, which implies some distinctive features of their decay mechanisms.

\begin{table*}[bth]
\caption{\label{tab_japforod}Experimental data of $\psp$ and $\jpsi$ decaying to decuplet-octet baryon final states. For branching fractions (${\cal B}_{\psp}\times 10^{6}/{\cal B}_{\jpsi}\times 10^{4}$), the first uncertainties are statistical, and the second are systematic. For the other quantities, the errors are merely statistical. The efficiency with star ($\ast$) is evaluated by virtue of the observed number of events $N^{obs}$, the total number of resonance events, and the corresponding branching fraction. }
\begin{ruledtabular}
\begin{tabular}{lllll}
  Mode  & $N^{obs}$       & Efficiency   & Branching Ratio           & Detector     \\
        &  (peak)         & (\%)         &($\times 10^{-6}/10^{-4}$) &     \\ \hline
\multicolumn{5}{c}{$\psp$ decay} \\ \hline
$\Xi(1530)^0 \overline{\Xi}^0$
  &$ 139.0 \pm 10.6$ &$5.85^{\ast}\pm 0.44$ &$5.3\pm 0.4\pm 0.3$ & BESIII~\cite{bes3xssaxsx21} \\
$\Xi(1530)^- \overline{\Xi}^+$
  &$ 199.5\pm 30.3$ &$6.36^{\ast}\pm 1.00$  &$7.0\pm1.1 \pm0.4$ & BESIII~\cite{zjybes3xxm} \\
        \hline
\multicolumn{5}{c}{$\jpsi$ decay} \\ \hline
$\Sigma(1385)^+ \overline{\Sigma}^-$
          &$77 \pm 9 $  &$2.63^{\ast}\pm 0.35$ &$3.4\pm 0.4\pm 0.8$ & DMII~\cite{dm2bbdkb} \\
          &$28 \pm 10$  &$6.8 \pm 2.4  $       &$3.1\pm 1.1\pm 1.1$ & MRKII~\cite{mrk2bbdk} \\
$\Sigma(1385)^- \overline{\Sigma}^+$
          &$74 \pm 8 $  &$2.87^{\ast}\pm 0.27$ &$3.0\pm 0.3\pm 0.8$ & DMII~\cite{dm2bbdkb} \\
          &$26 \pm 10$  &$6.7 \pm 2.5  $       &$2.9\pm 1.1\pm 1.0$ & MRKII~\cite{mrk2bbdk} \\
$\Xi(1530)^0 \overline{\Xi}^0$
          &$24 \pm 9 $  &$0.87^{\ast}\pm 0.33$ &$3.2\pm 1.2\pm 0.7$ & DMII~\cite{dm2bbdkb} \\
$\Xi(1530)^- \overline{\Xi}^+$
          &$75 \pm 11 $ &$1.48^{\ast}\pm 0.23$ &$5.9\pm 0.9\pm 1.2$ & DMII~\cite{dm2bbdkb} \\
        &$70186 \pm 544$ &$ 16.87\pm 0.11$  &$3.17\pm 0.02\pm 0.08$ & BESIII~\cite{bes3xsaxp20}  \\
$\Delta^+ \overline{p}$
          &$<50 $       &$-$                   &$<1.0~(90\%~CL)$    & DMII~\cite{dm2bbdkb} \\
$\Sigma(1385)^0 \overline{\Lambda}$
          &$<13 $       &$-$                   &$<2.0~(90\%~CL)$    & DMII~\cite{dm2bbdkb} \\
          &$<37 $       &$-$                   &$<0.082~(90\%~CL)$  & BESIII~\cite{bes3ssald13}  \\
\end{tabular}
\end{ruledtabular}
\end{table*}

\subsection{Decuplet-Octet mode}

The experimental data of $\psp$ and $\jpsi$ decay to decuplet-octet baryon final states are collected in Table~\ref{tab_japforod}.

For $\psp$ decay, the experimental information is too few to support further data analysis.

For $\jpsi$ decay, only the data that can be used to determine the branching fractions, are adopted. Moreover, For $\jpsi \to \Xi(1530)^- \overline{\Xi}^+$ final state, the branching fraction discrepancy between DM2 and BESIII are rather prominent. In order to keep the consistency of data, the number from BESIII is not used for parameters fitting. The fitting results are presented below:
\beq
\begin{array}{rcl}
  \phi  &=&-89.97^\circ\pm 37.17^\circ~, \\
     \Dp &=&~0.854 \pm 0.100~~, \\
      D  &=&~0.049 \pm 1.000~~, \\
  f_{mk2}&=&~0.890 \pm 1.031~~; \\
\end{array}
\label{fitjpsdom}
\eeq

and
\beq
\begin{array}{rcl}
  \phi  &=&+101.20^\circ\pm 71.87^\circ~, \\
     \Dp &=&~0.854 \pm 0.039~~, \\
      D  &=&~0.027 \pm 0.065~~, \\
  f_{mk2}&=&~0.890 \pm 0.242~~. \\
\end{array}
\label{fitjpsdop}
\eeq

It can be seen that the fitting error for the phase angle is fairly large. As a matter of fact, if the measurement of BESIII is included, we get the fitting phase angle $-77.94^\circ\pm 118.68^\circ~$ or       $+77.97^\circ\pm 349.72^\circ$, which means no reliable information about the phase angle could be extracted from the fit.

\begin{table*}[bth]
\caption{\label{tab_japforpp}Experimental data of $\psp$ and $\jpsi$ decaying to pseudoscalar meson pair final states. For branching fractions (${\cal B}_{\psp}\times 10^{5}/{\cal B}_{\jpsi}\times 10^{4}$), the first uncertainties are statistical, and the second are systematic. For the other quantities, the errors are merely statistical. The number of events due to the continuum process with star ($\ast$) are scaled to the resonance peak. Others are taken at $\sqrt{s}=3.67$ GeV for $\psp$ case and $\sqrt{s}=3.08$ GeV for $\jpsi$ case. The symbol ``$-$'' indicates that the continuum background is negligible.}
\begin{ruledtabular}
\begin{tabular}{llllll}
  Mode  & $N^{obs}$       & $N^{obs}$      & Efficiency   & Branching Ratio           & Detector     \\
        &  (peak)         &  (continuum)   & (\%)         &($\times 10^{-5}/10^{-4}$) &     \\ \hline
\multicolumn{6}{c}{$\psp$ decay} \\ \hline
$\pip\pim$ &$70.8  \pm 8.8$  &$40.4^{\ast} \pm 4.6$ &$16.4 \pm 5.4$
                                            &$0.76\pm 0.25\pm 0.06$ &CLEO-c~\cite{Metreveli12pp} \\
           &$11    \pm 3.32$ &$25.66  \pm 5.07$     &$16.7 \pm 16.7$
                                            &$0.8 \pm 0.8 \pm 0.2 $ &CLEO-c~\cite{Dobbs06pp} \\
$\kap\kam$ &$1431.3\pm 39.4$ &$106.9^{\ast}\pm 5.5$ &$72.4 \pm 2.2$
                                            &$7.48\pm 0.23\pm 0.39$ &CLEO-c~\cite{Metreveli12pp} \\
           &$157   \pm 12.53$&$68.20  \pm 8.26$     &$71.7 \pm 6.8$
                                            &$6.3 \pm 0.6 \pm 0.3 $ &CLEO-c~\cite{Dobbs06pp} \\
$\kskl$    &$478.0 \pm 23.0$ &$ - $                 &$37.0 \pm 1.8$
                                            &$5.28\pm 0.25\pm 0.34$ &CLEO-c~\cite{Metreveli12pp} \\
           &$53   \pm 7.28$  &$1.2   \pm 1.1$       &$29.5 \pm 4.1$
                                            &$5.8 \pm 0.8 \pm 0.4 $ &CLEO-c~\cite{Dobbs06pp} \\
           &$156  \pm 14$    &$- $                  &$21.5 \pm 1.9$
                                            &$5.24\pm 0.47 \pm 0.48$&BESII~\cite{bai04pp} \\
\hline
\multicolumn{6}{c}{$\jpsi$ decay} \\ \hline
$\pip\pim$  &$137.6 \pm 11.8$ &                    &$10.9 \pm 1.0$
                                             &$1.47\pm 0.13\pm 0.13$ & CLEO-c~\cite{Metreveli12pp} \\
            &$77.8 \pm 9.8$   &                    &$18.1 \pm 2.3  $
                                             &$1.58\pm 0.20\pm 0.15$ & MRKIII~\cite{mrk3pp85} \\
$\kap\kam$  &$1057.7\pm 32.8$ &                    &$43.1 \pm 1.4$
                                             &$2.86\pm 0.09\pm 0.19$ & CLEO-c~\cite{Metreveli12pp} \\
            &$107.0 \pm 10.7$ &                    &$16.5 \pm 1.7  $
                                             &$2.36\pm 0.24\pm 0.22$ & MRKIII~\cite{mrk3pp85} \\
$\kskl$     &$334.3 \pm 19.3$ &                    &$14.8 \pm 0.9$
                                             &$2.62\pm 0.15\pm 0.14$ & CLEO-c~\cite{Metreveli12pp} \\
            &$73.7 \pm 11.7$  &                    &$26.9 \pm 4.3  $
                                             &$1.01\pm 0.16\pm 0.09$ & MRKIII~\cite{mrk3pp85} \\
            &$2155 \pm 45$    & $- $         &$20.5 \pm 0.5$
                                             &$1.82\pm 0.04\pm 0.13$ & BESIII~\cite{bes2klks04} \\
    &$110203 \pm 504$  &$13 \pm 5 $   &$43.5 \pm 0.2$
                                             &$1.93\pm 0.01\pm 0.05$ & BESIII~\cite{bes3klks17} \\
\end{tabular}
\end{ruledtabular}
\end{table*}

\subsection{PP mode}

The total Born cross section of mesonic mode is similar to that of baryonic mode. As far as PP mode is concerned, it reads~~\cite{wymppdk}
\beq
\sigma_{B}(s) =\frac{4\pi \alpha^2}{3 s^{3/2}}
   |\ag(s)+\aga(s)+\ac(s)|^2~{\cal P}(s)~,
\label{bornxcpp}
\eeq
where the phase space factor ${\cal P}$ is expressed as
\beq
{\cal P} = \frac{2}{3s} \cdot q^3_f~,
~~ q^3_f =\sqrt{\frac{s}{4}-m^2}~.
\eeq
Here $\sqrt{s}$ is the center of mass energy and $m$ the mass of final state particle. The expressions for $\ac(s)$, $\aga(s)$, and $\ag(s)$ are the same as those shown in Eqs.~\eref{ampac}, \eref{ampap}, and \eref{ampag}.

The recent experimental data of $\psp$ and $\jpsi$ decay to pseudoscalar meson pair final states are summarized in Table~\ref{tab_japforpp}. Some data~\cite{Feldman:1977nj,Brandelik:1979hy,Vannucci:1977} taken more four decades ago are not included here. Neither adopted are the branching fractions measured by BaBar Collaboration~\cite{Lees15}. Since they used the initial state radiation technique, which is too different to be merged with other information. For $\psp$ decay, the numbers of events due to the continuum precess (as shown in Table~\ref{tab_japforpp} with $\ast$) in Ref.~\cite{Metreveli12pp} are scaled to the resonance peak, but there is a lack of the detailed information to recover their original appearance, so that these numbers can not be utilized in the following study.

It should be noted that for the data from CLEO~\cite{Dobbs06pp}, the number of events due to the continuum process is not subtracted from the signal but isolated as the continuum datum. The scaling factor $f_s$ is used to recover its original appearance.

For $\jpsi$ decay, the results in Ref.~\cite{Metreveli12pp} are obtained by cascade decay $\psp \to \pip\pim \jpsi$, then $\jpsi \to PP$. The feature of these data is totally distinctive from that of $\EE$ collider, therefore is not adopted in the fitting analysis.

The fitting procedure of $\psp$ and $\jpsi$ decay to pseudoscalar pair final states is exactly the same as that of $\psp$ and $\jpsi$ decay to two-body baryon final states. The performance of fitting yields
\beq
\begin{array}{rcl}
  \phi  &=&-58.19^\circ\pm 5.47^\circ~,\mbox{ or } ~+92.82^\circ\pm 5.62^\circ ; \\
     \Dp &=&~2.370 \pm 0.106~, \\
      D  &=&~0.831 \pm 0.054~, \mbox{ or } ~ 0.844 \pm 0.055~;\\
  f_{cleoa}&=&~1.034 \pm 0.103~, \\
  f_{cleob}&=&~0.977 \pm 0.113~, \mbox{ or } ~ 0.990 \pm 0.116~;
\end{array}
\label{fitpsppp}
\eeq
for $\psp$ decay and
\beq
\begin{array}{rcl}
  \phi  &=&-87.25^\circ\pm 8.60^\circ~,\mbox{ or } ~+92.14^\circ\pm 8.61^\circ ; \\
     \Dp &=&~1.211 \pm 0.003~, \\
      D  &=&~1.345 \pm 0.137~, \\
  f_{mk3}&=&~0.523 \pm 0.083~, \\
  f_{bes}&=&~0.947 \pm 0.020~;
\end{array}
\label{fitjpspp}
\eeq
for $\jpsi$ decay.

According to the fitting results, there is a large relative phase between the strong and electromagnetic amplitudes. Although the amplitudes of such final states are all due to $SU(3)$-breaking effect, the decay mechanisms of $\jpsi$ and $\psp$ are obviously different, as indicated through the ratio between $\Dp$ and $D$.

\subsection{Discussion}
For minimization, the MINUIT package, one of useful CERN packages in high energy physics
~\cite{minuit}, is utilized. The relevant information of fit is encapsulated in Talbe~\ref{fitinfmn}, including the values of chisquare, the fitting variables ($n_f$) composed of physical parameters and normalization factors, and the quantity of measured number of events ($N_D$). The minimization process is relatively insensitive to the initial trial parameters. When one set of solutions are obtained, the other can be easily acquired by just flipping the sign of phase angle, since usually only changed is the fitting value of the phase. Nevertheless, there is an exception for $\psp \to PP$ decay, some fitting values are distinguishable for the distinctive phase angle and also for the corresponding chisquare value as shown in Table~\ref{fitinfmn}.

\begin{table}[hbt]
\caption{\label{fitinfmn}The relevant information of fit, including the values of chisquare, the fitting variables ($n_f$) composed of physical parameters and normalization factors, and the quantity of measured number of events ($N_D$). The chisquare values correspond to the positive and negative (in parenthesis) phase are usually the same except for $\psp \to PP$ decay.}
\center
\begin{tabular}{llll}\hline \hline
Decay mode  & $\chi^2$  & $n_f$ & $N_D$  \\ \hline
$\psp \to B_{8}  \overline{B}_{8} $&  327.50  & 6+4  & 24   \\
$\jpsi\to B_{8}  \overline{B}_{8} $&  580.13  & 6+4  & 22   \\
$\psp \to B_{10} \overline{B}_{10}$&  65.28   & 4+3  & 8   \\
$\jpsi\to B_{10} \overline{B}_{10}$&  1.04    & 4+4  & 10   \\
$\jpsi\to B_{10} \overline{B}_{8} $&  17.01   & 3+1  & 6   \\
$\psp \to PP                           $&5.41(4.95)& 3+2  & 10   \\
$\jpsi\to PP                           $&  5.49    & 3+2  & 6   \\
\hline \hline
\end{tabular}
\end{table}

It is well known that in data analysis it is more convincing to use as few constraints as possible when fitting a data set, thus maximizing the number of degrees of freedom. For the analysis of this monograph, the degrees of freedom is equal to $N_D$ minus $n_f$, where $n_f$ is fixed by physical consideration and experimental characteristic, therefore, it is reasonable to include as much as possible experimental measurements. However, as disclosed in Table~\ref{fitinfmn}, the more data are included, the larger the chisquare values are.

From a pure viewpoint of hypothesis test~\cite{IGHuges2010,AGFrodeson19790}, the ratio of the chisquare value to the number of degrees of freedom should approximate one for a good fit, which is far from the case come across here. The discrepancies between the data and the proposed model are very unlikely to be due to the random statistical fluctuations, there may be some reasons for the disagreement. Firstly, in our data analysis, solely considered are the statistic uncertainties. If the systematic uncertainties are included as well, it is expected that the chisquare could be decreased to one half or one third of the present value.
Secondly, there could be some unknown systematic difference between different experimental measurements, as indicated by the measured branching fractions. Sometimes, the difference is really too great to be explained by statistical fluctuation. For example, for $\jpsi \to \Xi(1530)^- \overline{\Xi}^+$ decay, the branching fractions measured by DM2 and BESIII are respectively $5.9\pm 0.9\pm 1.2$ and $3.17\pm 0.02\pm 0.08$, the latter is almost a half of the former. Even more prominently, for $\jpsi \to \kskl$ decay, the branching fractions measured by MRKIII and CLEO-c are respectively $1.01\pm 0.16\pm 0.09$ and $2.62\pm 0.15\pm 0.14$, the former is almost one third of the latter. Therefore, the normalization factors are crucial for alleviating such a systematic difference. Anyway, such a kind of treatment may be not enough so that possibly produced are certain great deviations between data and expected evaluations.
Thirdly, the merit of the analysis approached adopted in this monograph is that the experimental information has been fully utilized. However, sometimes the details of experiments can not be fully available, which will degrade the validity of analysis.
Fourthly, it also exists the possibility that the present parametrization form is not exquisite enough to describe all data perfectly, say, in Ref.~\cite{Baldini19} authors introduce more parameters to describe  the $\jpsi$ decaying to octet baryon pair final states. However, further more precise and consistent experimental data are need to furnish quantitative evidence for or against the present phenomenology model.

If scrutinizing the fitting results listed in \eref{fitpspoo}, \eref{fitjpsioo}, \eref{fitpspdd}, \eref{fitjpsidd}, \eref{fitjpsdom}, \eref{fitjpsdop}, \eref{fitpsppp}, and \eref{fitjpspp}, it is generally in line with physical expectation. The relative strengthen of $SU(3)$-conserving effect (denoted by $A$) is much greater than that of $SU(3)$-breaking effect (denoted by $D,~F,~\Dp,$ and $\Fp$). For decuplet-octet mode, since only $SU(3)$-breaking effect plays a leading role, the branching fractions are generally one order of magnitude lower than those of octet-octet and decuplet-decuplet modes. The same is also true for pseudoscalar-pseudoscalar mode. As a conclusion, we accept the present fitting results as reasonable ones. It also indicates that much more systematic and accurate experimental measurements will play a vital role to clarify the qualm issues at present.

\subsection{Comment}
The old-version of parametrization, appearing as sum rules of coupling-constant, can retrospect to the sixty of twentieth century~\cite{Muraskin1963,Gupta1964a,Gupta1964b}. The notion of a spurion octet for $SU(3)$ symmetry breaking effect was already adopted.

A new-version of parametrization appeared in the seventy of twentieth century~\cite{Kowalski:1976mc}. Both strong and electromagnetic breaking effects are taken into account and parametrization forms for $\jpsi$ decaying to VP final states and octet baryon pair final states are provided. The Refs.~\cite{zmy2015,Baldini19,moxh2022} focus on the parametrization of baryon pair final states. The more general result is acquired in Ref.~\cite{Haber}, in which the concepts of both flavor-$SU(3)$ singlet and spurion octet are combined to give the systematic parametrization for $\jpsi$ and $\eta_c$ decays to two- and three-meson final states. Afterwards, the parametrization including higher order effect due to symmetry breaking effect is considered as well~\cite{Seiden88,Morisita:1990cg}.

In principle, all achievements of new-version parametrization can be subsumed in the scenario proposed in this mongraph, which supplies a unified foundation for parametrization of charmonium decay. Moreover, the definite and comparative concise forms of parametrization are amenable to experimental verification.

As far as measurement of phase angle is concerned, the most model-independent approach is through energy scan. A recent work~\cite{bes3fsofjpsi19} at BESIII is performed by using 16 energy points of $\EE$ annihilation data collected in the vicinity of $\jpsi$ resonance. The relative phase between strong and electromagnetic amplitudes is measured to be $(+84.9 \pm 3.6)^\circ$ or $(-84.7 \pm 3.1)^\circ$. In addition, the cross section line-shapes of $\jpsi \to \MM$ and $\jpsi \to \eta \pp$ with $\eta \to \pp\piz$ are also investigated. Measured is the relative phase between $\jpsi$ resonance and continuum decays, which is consistent with zero within fitting uncertainty. Nevertheless, the demerit of this kind of analysis leads us to lose the elaborate insight into the details of decay mechanism, especially the information about $SU(3)$-conserving and $SU(3)$-breaking effects.

As last, a few words about the multiple solution issue. When fitting cross sections with several resonances or interfering background and resonances, one usually obtains multiple solutions of parameters with equal fitting quality. Such a phenomenon was firstly noticed experimentally~\cite{moxh2010prd,yuancz2010ijmpa}, then some studies are performed from a mathematical point of view~\cite{yuancz2011cpc,zhuk2011ijmpa,hanx2018cpc,baiyu2019prd}. Especially in Ref.~\cite{baiyu2019prd}, the source of multiple solutions for a combination of several resonances or interfering background and resonances is found by analyzing the mathematical structure of the Breit-Wigner function. It is proved that there are exactly $2^{n-1}$ fitting solutions with equal quality for $n$ resonances, and the multiplicity of the interfering background and resonances depends on zeros of the amplitudes in the complex plane. For our studies, the interference between resonance and continuum amplitudes leads to just two solutions of phase angle.

\section{Summary}\label{xct_sum}
Based on the flavor-singlet principle, assuming the flavor symmetry breaking effects (both strong and electromagnetic breaking effects) as a special $SU(3)$ octet, the effective interaction Hamiltonian can be obtained in tensor form for all kinds of two-body final states decaying from a charmonium resonance. It is the first time that such a scheme is acquired to systematically parameterize various kinds of baryon and meson pair final states in the light of a single and simple principle. Such a uniform parametrization scheme of  charmonium decay modes facilitates the study of the relative phase between the strong and electromagnetic amplitudes. In data analysis of samples taken in $\EE$ collider, the details of experimental effects, such as energy spread and initial state radiative correction are taken into consideration in order to make full advantage of experimental information and acquire the comprehensive results.

Based on fitting results, on one hand it indicates that there exists a large relative phase around $90^\circ$ between the strong and electromagnetic amplitudes; on the other hand, comparing parameters of $\jpsi$ and $\psp$ decays, the pattern of relative strength is similar but is not exactly the same, which implies some distinctive decay mechanisms of two resonances. Such a study makes it urgent that further more precisely and systematically experimental measurements should be performed based on BESIII colossal data sample of charmonium decay, in order to disclose more subtle feature of decay mechanisms beyond the prescription of symmetry analysis.

By virtue of present analysis, the uniform parametrization scheme provides a general description for charmonium two-body decays and lays a basis for more profound dynamics exploration in the future.

\section*{Acknowledgment}
This work is supported in part by National Key Research and Development Program of China under Contracts Nos.~2020YFA0406302 and 2020YFA0406403, and by IHEP funding under Contract No. E25471JY10.

\end{document}